\documentclass{aa}
\usepackage{graphicx}
\usepackage{epsf}
 \def\sun{\odot}
 \def\ms{M_{\sun}}
 \def\ie{{\it i.e.,}}	
 \def\eg{e.g.,}	

\begin{document}
\title{Stability analysis of relativistic jets from collapsars and 
       its implications on the short-term variability of gamma-ray bursts}
\author{M.-A.Aloy$^{1)}$, J.-M.Ibanez$^{2)}$, J.-A.Miralles$^{3)}$,
        V.Urpin$^{2,4)}$}
\offprints{M.-A.Aloy, \\ e-mail: maa@mpa-garching.mpg.de}
\institute{$^{1)}$ Max-Planck-Institut f\"{u}r Astrophysik,
                   Karl-Schwarzschild-Str. 1, Postfach 1523,
                   Garching, D-85748, Germany \\
           $^{2)}$ Departamento de Astronom\'{\i}a y Astrof\'{\i}sica, 
                   Universidad de Valencia, E-46100 Burjassot, Spain \\
           $^{3)}$ Departament de F\'{\i}sica Aplicada, Universitat d'Alacant,
                   Ap. Correus 99, E-03080 Alacant, Spain \\
           $^{4)}$ A.F.Ioffe Institute of Physics and Technology,
                   194021 St.Petersburg, Russia \\}
\date{Received 2 July 2002 / Accepted 30 August 2002}

\abstract{We consider the transverse structure and stability
properties of relativistic jets formed in the course of the collapse
of a massive progenitor. Our numerical simulations show the presence
of a strong shear in the bulk velocity of such jets. This shear can be
responsible for a very rapid shear--driven instability that arises for
any velocity profile. This conclusion has been confirmed both by
numerical simulations and theoretical analysis. The instability leads
to rapid fluctuations of the main hydrodynamical parameters (density,
pressure, Lorentz factor, etc.). However, the perturbations of the
density are effectively decoupled from those of the pressure because
the beam of the jet is radiation--dominated. The characteristic growth
time of instability is much shorter than the life time of the jet and,
therefore, may lead to a complete turbulent beam.  In the course
of the non-linear evolution, these fluctuations may yield to internal
shocks which can be randomly distributed in the jet. In the case that
internal shocks in a ultrarelativistic outflow are responsible for
the observed phenomenology of gamma-ray bursts, the proposed
instability can well account for the short-term variability of
gamma-ray light curves down to milliseconds.

}

\authorrunning{Aloy et al.}
\titlerunning{Stability analysis of jets from collapsars}
\maketitle

\section{Introduction}

Catastrophic stellar events like massive stellar core collapse (see,
e.g., van Putten \cite{Putten01}, M\'esz\'aros \cite{Meszaros01} and
references therein) or merging of a binary neutron star (Paszynski
\cite{Paszynski86}, Eichler et al. \cite{Eichleretal89}, Mochkovitch
et al. \cite{Mochkovitch93}) have been proposed to explain the
energetics of gamma-ray bursts (GRB).  Nowadays, an increasing amount
of observational evidence stresses the association, at least, of some
GRB events with massive progenitors (Bulik, Belczynski \& Zbijewski
\cite{BBZ99}; Bloom et al. \cite{Bloometal99}; Hanlon et
al. \cite{Hanlonetal00}; Reeves et al. \cite{Reevesetal02}).  One of
the most attractive scenarios is the collapsar model (Woosley
\cite{Woosley93}, MacFadyen \& Woosley \cite{MW99}) which can produce
the required energy of a GRB even at cosmological distances. This
model involves the collapse of the central core of a massive, evolved
star to a newly-formed black hole (BH).  The progenitor star can be,
for instance, a rotating Wolf-Rayet star
(\cite{Woosley93}). Hydrodynamic simulations of collapsars have been
performed for a 35$M_{\odot}$ main-sequence star whose 14$M_{\odot}$
helium core collapses to form 2-4$M_{\odot}$ black hole. Provided that
the core has a sufficient amount of angular momentum, a massive
geometrically thick accretion disc of several tenths of a solar mass
can be formed around the BH. The BH accretes matter from the disc at
rates of the order of solar masses per second. The burst is generated
by a local energy deposition due to the annihilation of
$\nu-\bar{\nu}$ coming from the accretion disk or/and due to the
release of the BH spin energy by means of magnetic fields. In either
case, the energy is preferentially released along the rotation axis,
close to the central engine and gives rise to a relativistically
expanding bubble of radiation and pairs. The duration of the burst is
given by the hydrodynamic time scale for the core of the star and its
Helium envelope to collapse or by the viscous evolution time for the
accretion disc, whichever is greater. In order to produce a
long-duration GRB with a complex pulse structure, accretion is argued
to proceed over a period of time comparable to the prompt phase of a
GRB. The energy released due to accretion is sufficient to drive a
collimated, baryon-dilute fireball that penetrates the outer layers of
the star forming a relativistic jet (Aloy et al. \cite{Aetal00}).

Recent observations indicate that the rapid temporal decay of several
GRB afterglows is consistent with the evolution of a highly
relativistic jet after it slows down and spreads laterally rather than
with a spherical blast wave (Sari, Piran \& Halpern \cite{SPH99},
Halpern et al. \cite{Halpernetal99}, Kulkarni at
al. \cite{Kulkarnietal99}, Rhoads \cite{Rhoads99}). Therefore,
formation of relativistic jets of baryon-clean material with bulk
Lorentz factors $\sim 10^{2}- 10^{3}$ represents a mayor problem in
the collapsar model of GRB. Detailed simulations of formation and
propagation of a relativistic jet in collapsars have been performed by
Aloy et al. (\cite{Aetal00}). Assuming an enhanced efficiency of
energy deposition in polar regions (with a constant or varying
deposition rate), the authors obtained an ultrarelativistic jet along
the rotation axis, which is highly focused (with the half-opening
angle $\sim 6-10^{\circ}$) and is capable of penetrating the star. The
simulations were performed with the multidimensional relativistic
hydrodynamic code GENESIS (Aloy et al. \cite{Aetal99}) using a
two-dimensional spherical grid. A relativistic jet in this model forms
within a fraction of a second and exhibits the main morphological
elements of ``standard'' jets: a terminal bow shock, a narrow cocoon,
a contact discontinuity, and a hot spot. The maximum Lorentz factor is
as large as $\sim 25-30$ at break-out, and can be even greater after
the break-out ($\approx 44$ at the end of simulations). This Lorentz
factor is only a bit smaller than the critical value $\sim
10^{2}-10^{3}$ that one requires for the fireball model (Cavallo \&
Rees \cite{CR78}, Piran \cite{Piran99}).
  
According to current views, the GRB is made either as the jet
encounters a sufficient amount of mass in circumstellar matter or by
internal shocks in the jet (Rees \& M\'esz\'aros \cite{RM92},
\cite{RM94}, Daigne \& Mockkovitch \cite{DM98},
\cite{DM00}). Therefore, the properties of jets from collapsars are of
crucial importance for understanding the mechanism of GRBs. For
example, the non-uniformity across the jet can influence the structure
of internal shocks and, hence, the gamma-ray emission from these
shocks. A transverse gradient can also drastically change the temporal
decay rate of afterglows, making the decay flatter or steeper depending
on the transverse structure (M\'esz\'aros, Rees \& Wijers
\cite{MRW98}). Gamma-ray light curves also show a great
diversity of time dependences ranging from a smooth rise and
quasi-exponential decay, through light curves with several peaks, to
variable light curves with many peaks, and substructure sometimes down
to milliseconds. Hydrodynamic instabilities arising in the jet can be
responsible for the fine time structure of GRBs in a row with a
temporal variability caused by accretion onto the BH and the
circumstellar interaction. This complex time dependence can provide
clues for the understanding of the geometry and physics of the
emitting regions. Apart from this, instabilities and their associated
fluctuating hydrodynamic motions can lead to the generation of
magnetic fields in a highly conductive $e^{-} e^{+}$ plasma. In the
presence of turbulent magnetic fields, the electrons can produce a
synchrotron radiation spectrum (M\'esz\'aros \& Rees \cite{MR93}, Rees
\& M\'esz\'aros \cite{RM94}) similar to that observed (Band et
al. \cite{Bandetal93}). The inverse Compton scattering of these
synchrotron photons can extend the spectrum even into the GeV range
(M\'esz\'aros, Rees \& Papathanasiou \cite{MRP94}).

In the present paper, we consider the stability properties of jets
which are formed and propagate in collapsars. Large--scale hydrodynamic
instabilities can be the reason for the observed morphological
complexity of ``standard'' jets, and this has motivated many
analytical (see, e.g., Birkinshaw \cite{Birkinshaw84},
\cite{Birkinshaw91}, \cite{Birkinshaw97}, Hardee \& Norman
\cite{HN88}, Zhao et al. \cite{Zhaoetal92}, Hanasz \& Sol \cite{HS96})
and numerical (Hardee et al. \cite{Hardeeetal92}, Hardee et
al. \cite{Hardeeetal98}, Bodo et al. \cite{Bodoetal98}, Micono et
al. \cite{Miconoetal00}, Agudo et al. \cite{Agudoetal01}) studies of
the stability properties of jets.  Usually, the jet is considered as a
beam of gas with one bulk velocity and constant density surrounded by a
very narrow shear layer separating it from the external medium. One
possible mechanism of destabilizing such jets is often attributed to
the well-known Kelvin-Helmholtz instability which, in its classical
formulation, is the instability of a tangential discontinuity between
two flows, generally having different density (see, e.g., Landau \&
Lifshitz \cite{LL78}, Chandrasekhar \cite{Chandrasekhar81}). However,
jets in collapsars show a complex structure with the presence of
strong transverse shear and substantial density stratification (Aloy
et al. \cite{Aetal00}). Obviously, the stability properties of such
sheared, stratified jets may well be different from those of jets with
constant bulk velocity and density.

The outline of this paper is as follows. In Section 2, we represent
the results of numerical calculations and discuss the transverse
structure of a jet originating in a collapsar. In Section 3, we represent
the linear stability analysis of a sheared, stratified jet by making
use of a WKB-approximation. Finally, in Section 4, we discuss the 
possible role of the instability in the evolution of jets.

\section{Numerical simulations of jets from collapsars}
\label{numsim}

The simulations performed are the same as in Aloy et
al. (\cite{Aetal00}). Here we reproduce the technical details for
completeness. The computational mesh is a two-dimensional spherical
grid (with coordinates $r, \theta$).  In the $r$--direction the
computational grid has 200 zones spaced logarithmically between the
inner boundary and the surface of the helium star at $R = 2.98 \times
10^{10}\,$cm. Equatorial symmetry is assumed and the working angular
resolution is $0.5^{\circ}$ close to the polar region ($0^{\circ} \le
\theta \le 30^{\circ}$) and decreases logarithmically between
$30^{\circ} \le \theta \le 90^{\circ}$.

A central Schwarzschild BH of mass $3.762\ms$ provides the
gravitational field that couples the system. Self-gravity of
the star is neglected, \ie we consider only the gravitational
potential of the BH.  Our equation of state (EoS) includes the
contributions of non-relativistic nucleons treated as a mixture of
Boltzmann gases, radiation, and an approximate correction due to
$e^+e^-$--pairs as described in Witti, Janka \& Takahashi (\cite{WJT94}).
Complete ionization is assumed, and the effects due to degeneracy are
neglected. We advect nine non-reacting nuclear species which are
present in the initial model: C$^{12}$, O$^{16}$, Ne$^{20}$,
Mg$^{24}$, Si$^{28}$, Ni$^{56}$, He$^{4}$, neutrons and protons.

In a consistent collapsar model a jet will be launched by any process
which gives rise to a local deposition of energy and/or momentum, as
\eg  $\nu \bar\nu$--annihilation, or magneto-hydrodynamic processes. We
reproduce such a process by depositing energy at a prescribed rate
homogeneously within a $30^{\circ}$ cone around the rotation axis.  In
the radial direction the deposition region extends from 200\,km to
600\,km.  We have investigated a constant energy deposition rate $\dot
E = 10^{50}\,$erg\,s$^{-1}$, of $\dot E = 10^{51}\,$erg\,s$^{-1}$, and
a varying deposition rate with a mean value of $10^{50}\,$erg\,s$^{-1}$. 
The constant rates bracket the expected deposition rates of collapsar 
models, while the varying rate mimics, \eg time-dependent mass accretion 
rates resulting in time-dependent $\nu \bar \nu $--annihilation 
(MacFadyen \& Woosley \cite{MW99}).

We have endowed the star with a Gaussian atmosphere, which at $R_{\rm
a} = 1.8\, R$ passes over into an external uniform medium with a
density $10^{-5}$\,g\,cm$^{-3}$ and a pressure $10^{-8} p(R)$. The
computational domain is extended to $R_t = 2.54 R$ with 70 additional
zones. At the end of the simulation, a relativistic jet has broken out
from the surface of the star and propagates through the added
atmosphere.

\begin{figure*}
\epsfxsize=15cm
\epsffile{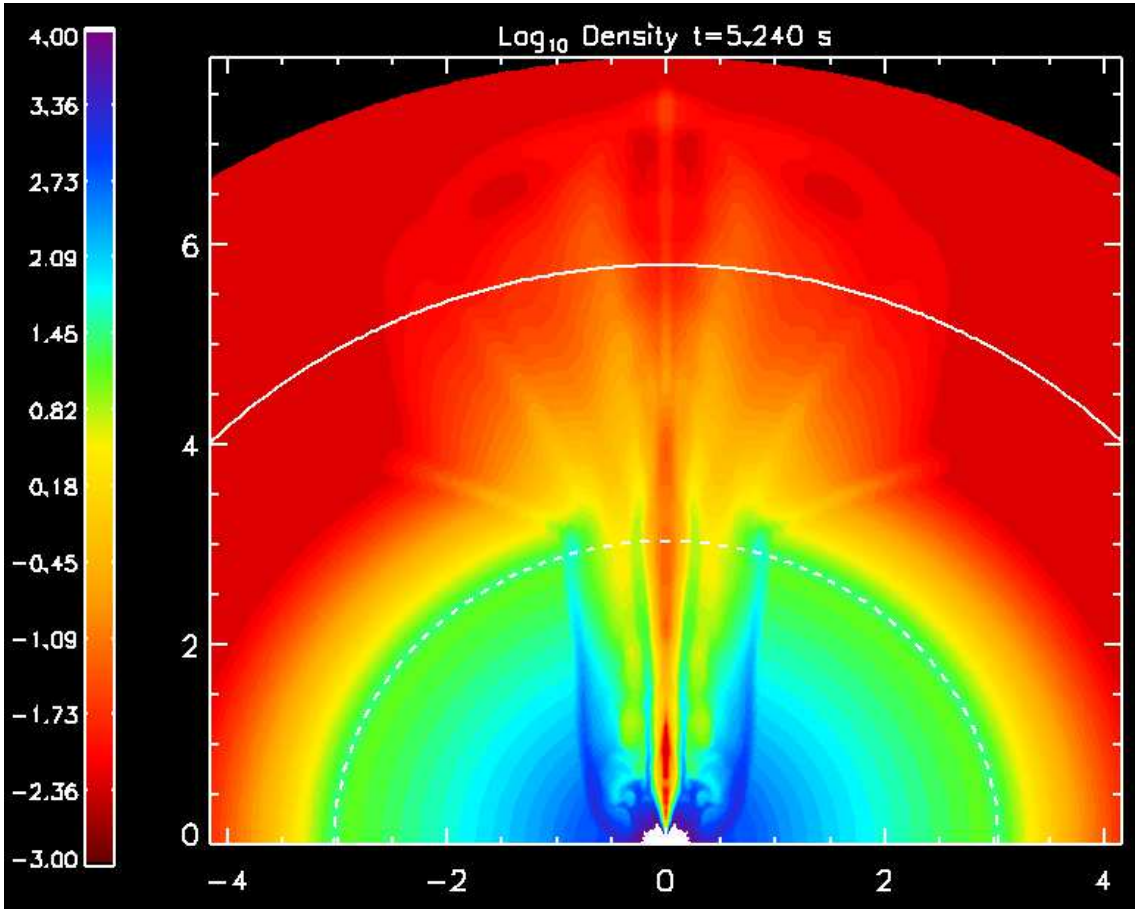}
\caption{Snapshot of the logarithm of the rest--mass density
after an evolution time of  $5.24\,$s. The color scale indicates the
value of the logarithm of the rest--mass density (in CGS units). The
solid arc marks the limit of the exponentially decaying atmosphere
while the dashed one marks the surface of the collapsing star. The
numbers around the box represent distances in units of $10^{10}\,$cm.}
\label{fig:isodensity}
\end{figure*}

\begin{figure*}
\epsfxsize=15cm
\epsffile{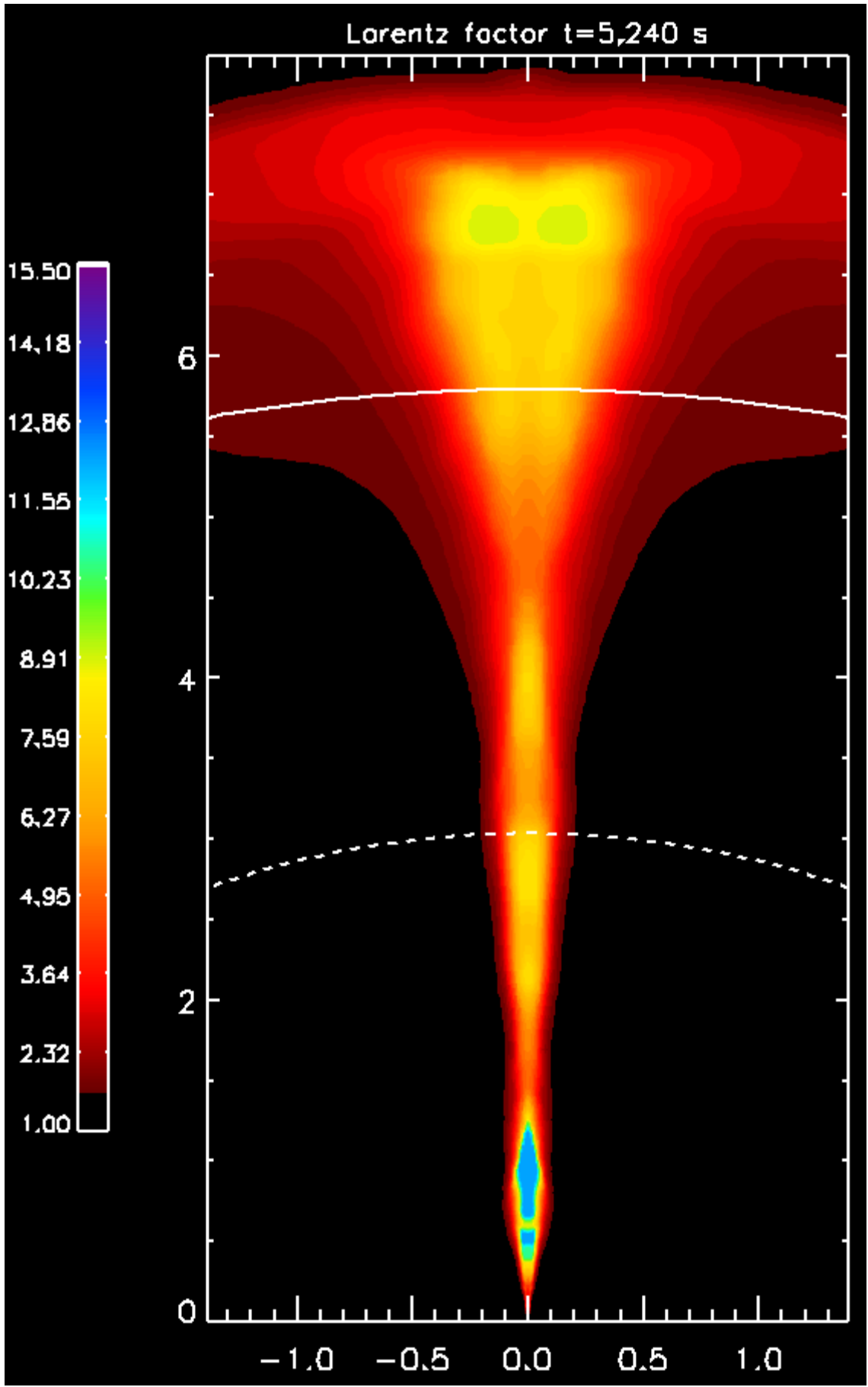}
\caption{Snapshot of the Lorentz factor after an evolution time of
$5.24\,$s. The color scale indicates the value of the Lorentz factor
at every point of our computational domain. The solid arc marks the limit of
the exponentially decaying atmosphere while the dashed one marks the
surface of the collapsing star. The number around the box represent
distances in units of $10^{10}\,$cm. Note that the horizontal scale
has been enlarged by a factor of 3 in order to display more details of
the stratified structure of the fluid in the jet.}
\label{fig:isolorentz}
\end{figure*}

We will focus in the following on the model C50 of Aloy et
al. (\cite{Aetal00}). The logarithm of the rest-mass density and the
Lorentz factor after an evolution time of $5.240\,$s are displayed in
Figs.~\ref{fig:isodensity} and \ref{fig:isolorentz}. Note that we have
enlarged the horizontal scale in Fig.~\ref{fig:isolorentz} in order to
improve the readability of the transverse structure of the jet.

In Fig.~\ref{fig:lorentz}, we plot the radial dependence of a Lorentz
factor, $\Gamma$, for a few selected polar angles. This dependence
shows quite well that the jet is highly collimated. The Lorentz factor
reaches maximum values at the axis where it can be as large as $\sim
20-30$. However, $\Gamma$ drops very rapidly with increasing polar
angle and, at $\theta = 10^{\circ}$, it becomes close to 1 in the main
fraction of a collapsar volume ($r \geq 10^{8}$cm). Within the
collapsar, on average, the Lorentz factor increases with $r$ because
of the decreasing density. The maximum in $\Gamma$ is caused by a very
strong rarefaction wave behind the recollimation shock. Note that the
Lorentz factor shows a well developed small--scale structure in the
region with $r \geq 10^{8}$cm. The characteristic length scale of
variations increases from $\sim 10^{7}$cm at $r=10^{8}$cm to $\sim
10^{9}$cm at $r=10^{10}$cm. These fluctuations may indicate the
presence of a very rapid hydrodynamic instability in the jet.

\begin{figure*}
\epsfxsize=15cm
\epsffile{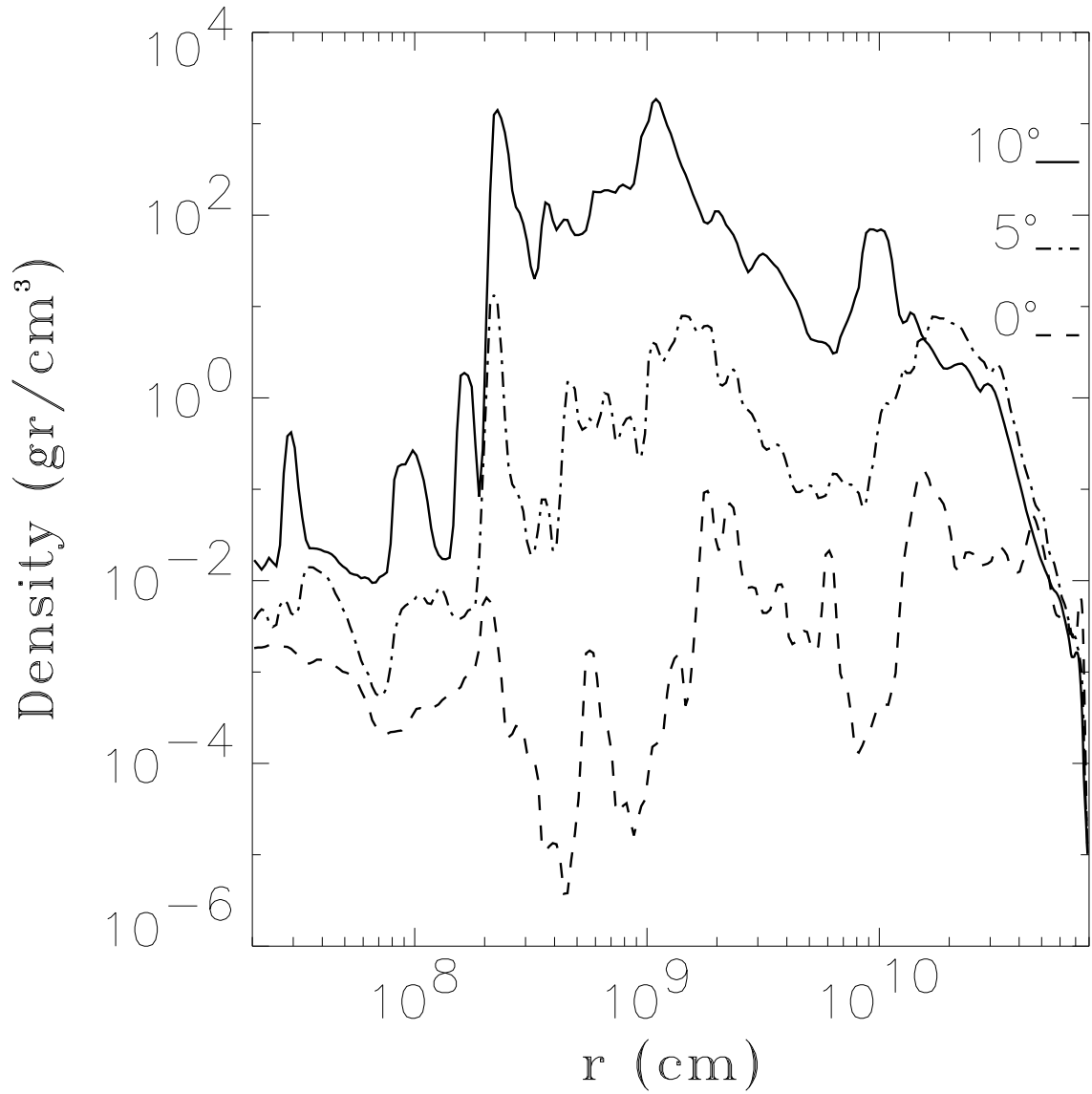}
\caption{Rest--mass density as a function of the radius for three
different angles ($\theta\approx 0^{\circ}, 5^{\circ}, 10^{\circ}$;
see legends in the figure) after an evolution time of $5.24\,$s.}
\label{fig:density}
\end{figure*}

Fig.~\ref{fig:density} shows the radial dependence of the rest-mass
density for a few selected polar angles. The density shows an even
stronger dependence on $\theta$ than does the Lorentz factor. For
instance, the density contrast between the axis (where the density is
extremely low, $\sim 10^{-3} - 10^{-4}$g/cm$^{3}$) and $\theta =
10^{\circ}$ is 4-6 orders of magnitude. This is in large contrast to
the relatively small average change of the density with $r$ which does
not exceed one order of magnitude when $r$ varies from $2 \times
10^{8}$cm to $R$. The main reason for this change is the fact that the
radial grid stretches logarithmically and, therefore, as $r$ grows,
there is a loss of numerical resolution that may suppress small--scale
structures. Like the Lorentz factor, the density shows a very
remarkable small--scale structure approximately with the same length
scale as $\Gamma$. However, fluctuations of the density are larger (up
to a factor of $10^{2}-10^{3}$).  The reason for this behavior of the
density is that the beam is radiation dominated according to our EoS
(see Sect.~\ref{numsim}) because the average temperature of the beam
is $\approx 5\times10^8\,$K. In this radiation--dominated flow, the
relative variation of the density, resulting from variations of
temperature and of pressure, can be written as
\begin{eqnarray}
\frac{\Delta \rho}{\rho} \approx 
	\frac{p}{\rho K  T}  ( \frac{\Delta p}{p} - 4 \frac{\Delta T}{T}),
\end{eqnarray}
where $T$, $p$ and $K$ are the temperature, the pressure and the
Boltzmann constant, respectively. As the relative variations of the
pressure and of the temperature are similar, it turns out that
\begin{eqnarray}
\frac{\Delta \rho}{\rho} \sim \frac{p}{\rho K  T} \frac{\Delta p}{p}. 
\end{eqnarray}
Therefore, $p / (\rho K T) >> 1$, we conclude that the relative
variations of the density must be much larger than the relative
variations of pressure or of temperature in a radiation--dominated
beam.

\begin{figure*}
\epsfxsize=15cm
\epsffile{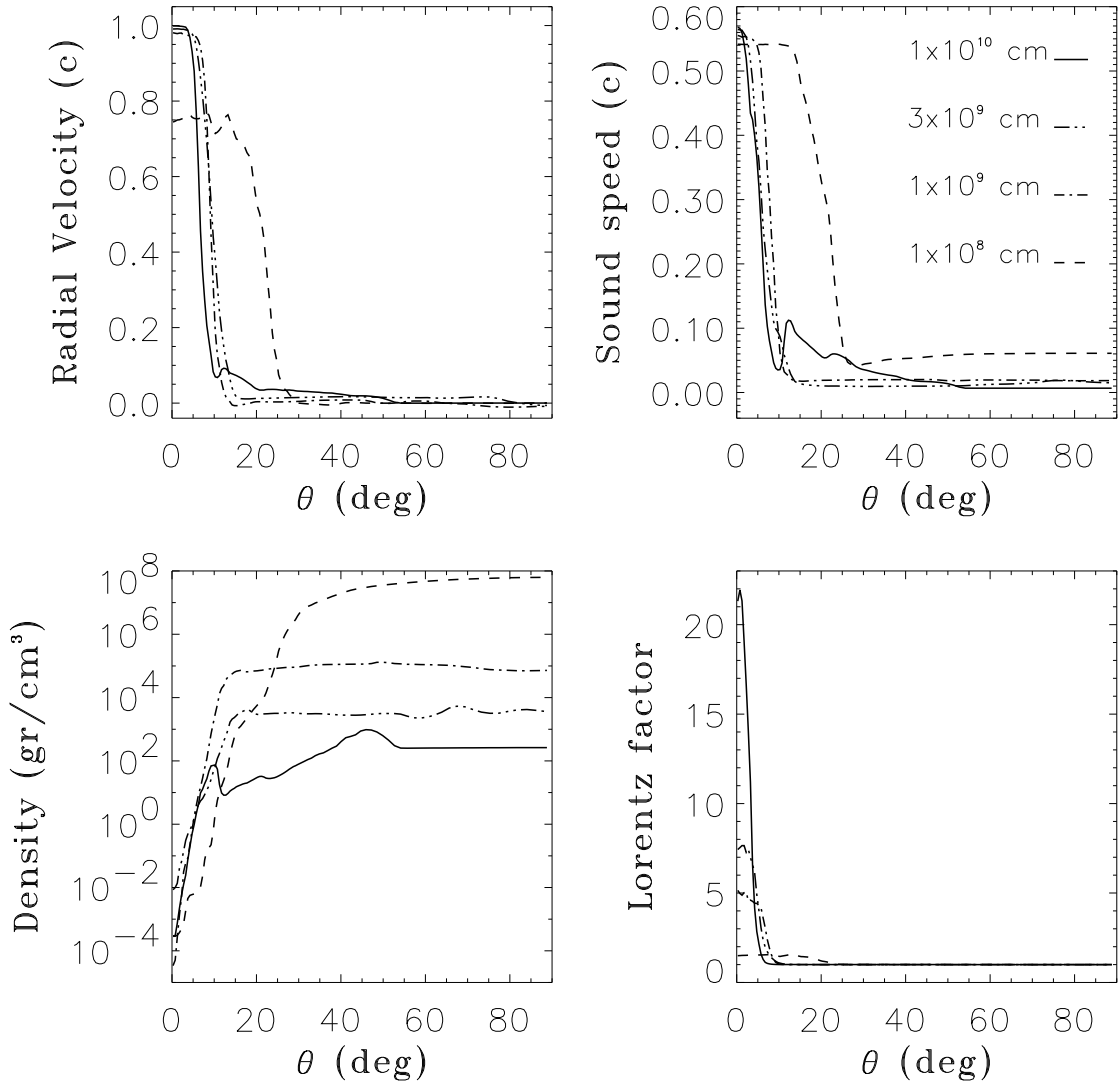}
\caption{Plots of four physical quantities as a function of the polar
angle, $\theta$, at four different distances ($10^8, 10^9,
3\times10^9$ and $10^{10}$ cm; see legends in the figure). The upper
left panel corresponds to the radial velocity (in units of $c$); the
lower left panel to the rest--mass density and, the upper and lower
right panels correspond to the sound speed (in units of $c$) and to
the Lorentz factor, respectively. The evolution time is $5.24\,$s.}
\label{fig:polarplots}
\end{figure*}

In Fig.~\ref{fig:polarplots}, we show transverse dependences of the
radial velocity, sound speed, density and Lorentz factor at different
distances.  All quantities are collimated in a cone with a
half-opening angle of about 6-10$^{\circ}$ with a better collimation
for the Lorentz factor and a worse one for the density. Only close to
the black hole ($r \sim 10^{8}$ cm) is the collimation of the
represented quantities (except the Lorentz factor) worse, $\sim
20^{\circ}$. Note also that all quantities except the Lorentz factor
exhibit almost a self-similar behavior at $r > 10^{8}$cm.  The
velocity profiles show the presence of a strong shear which is crucial
for the dynamics of jets and their stability properties. The sound
speed is close to the maximum possible value, $c/ \sqrt{3}$, in a
significant fraction of the jet volume, \ie the jet plasma is
relativistic in the sense of both the bulk velocity and the thermal
content. The mean temperature is $\sim 5 \times 10^{8}$K, and the
pressure is dominated by radiation. That is why the sound speed is
relativistic despite the temperature being non-relativistic for
particles. Note the presence of a substantial transverse gradient of
the sound speed that may play an important role in the formation of
internal shocks. Jets in collapsars exhibit a highly inhomogeneous
density in contrast to the standard jet models. The density becomes
extremely low at the jet axis and it grows almost exponentially with
the angle reaching larger values at the jet surface.  This behavior is
caused by a velocity profile decreasing with $\theta$ and, likely, is
typical for all sheared jets.

\section{The dispersion equation for perturbations}

We model the jet from collapsars by an infinitely long plasma cone
with a half-opening angle $\theta_{0}$. From simulations, the typical
value of $\theta_{0}$ is somewhat around 6-10$^{\circ}$. Plasma inside
the jet moves with a velocity ${\bf V} = V(r, \theta) {\bf e}_{r}$
with respect to the ambient medium; $r$, $\theta$, $\varphi$ are the
spherical coordinates with ${\bf e}_{r}$, ${\bf e}_{\theta}$, ${\bf
e}_{\varphi}$ being the corresponding unit vectors. As seen from the
behavior of the velocity profile (Fig.~\ref{fig:polarplots}), we can
distinguish two different regions within the jet: the central core
where the flow is ultrarelativistic, and the surrounding transition
layer where the velocity decreases from relativistic to
non-relativistic values. In the core region, the Lorentz factor is
large, $\geq 3$, and may even reach values as high as $\sim 25-30$
(see Fig.~\ref{fig:lorentz}). It is convenient to distinguish between
these two regions in our analysis because of the difficulty of
describing both relativistic and non-relativistic flows within the
framework of the same analytical formalism. Basically, the thickness
of the transition layer is smaller than the radius of the core. Since
our main interest is the processes in the core region we will model
the effect of the transition layer in terms of the boundary
conditions. Note that outside this transition layer there exists an
extended region where the flow is non-relativistic but we neglect the
influence of this region because of its low energy.

In relativistic hydrodynamics the continuity, momentum, and energy 
equations read (see, e.g., Weinberg \cite{Weinberg72})
\begin{equation}
\Gamma \left( \frac{\partial \rho}{\partial t} + {\bf v}
\cdot \nabla \rho \right) + \left( \rho + \frac{p}{c^{2}}
\right) \left[ \frac{\partial \Gamma}{\partial t} +
\nabla \cdot ( \Gamma {\bf v} ) \right] = 0,
\label{eq:continuity}
\end{equation}
\begin{equation}
\Gamma^{2} \left( \rho + \frac{p}{c^{2}} \right) \left[
\frac{\partial {\bf v}}{\partial t} + ({\bf v} \cdot \nabla )
{\bf v} \right] = - \nabla p 
- \frac{{\bf v}}{c^{2}} \frac{\partial p}{\partial t} ,
\label{eq:momentum}
\end{equation}
\begin{equation}
\frac{\partial}{\partial t} (p n^{- \gamma}) + {\bf v} \cdot
\nabla (p n^{- \gamma}) = 0,
\label{eq:energy}
\end{equation}
where $\Gamma = (1 - v^{2}/c^{2})^{-1/2}$ is the Lorentz factor,
$p$ and $\rho$ are the gas pressure and density, respectively; $n$ 
is the number density in the fluid's rest frame; $\gamma$ is the 
adiabatic index.  

Our analysis of stability is based on the linearized set of equations
(\ref{eq:continuity})-(\ref{eq:energy}). Small perturbations will be
marked by the index 1; for unperturbed quantities subscripts will be
omitted with the exception of vector components. We consider the
instability which arises on a time scale much shorter than the
characteristic evolution time scale of the jet. Therefore, we can
adopt a quasi-steady approximation neglecting the time dependence of
unperturbed quantities. In this approximation, the linearized
continuity, momentum and energy equations are
\begin{eqnarray}
\Gamma \left( \dot{\rho_{1}} + 
{\bf v}_{1} \cdot \nabla \rho + {\bf V} \cdot \nabla \rho_{1} \right) + 
\rho_{*} \left[ \nabla \cdot ( \Gamma {\bf v}_{1}) + 
\right.
\nonumber \\
\left.
\frac{\Gamma^{3} V}{c^{2}} \left( \dot{v}_{1r} + 
V \frac{\partial v_{1r}}{\partial r} \right) \right]
= - \frac{v_{1r}}{c^{2} r^{2}} \frac{\partial}{\partial r} (r^{2}
\rho_{*} V^{2} \Gamma^{3}) 
\nonumber \\
- \frac{1}{r^{2}} \left(\rho_{1} + \frac{p_{1}}{c^{2}} \right)
\frac{\partial}{\partial r} (r^{2} \Gamma V),
\label{eq:continuity2}
\end{eqnarray}
\begin{eqnarray}
\Gamma^{2} \rho_{*} \left[ \dot{\bf v}_{1} + ({\bf v}_{1} \cdot \nabla) 
{\bf V} + ({\bf V} \cdot \nabla) {\bf v}_{1} \right] + \nabla p_{1}
+ \frac{{\bf V}}{c^{2}} \dot{p}_{1} =
\nonumber \\
\frac{{\bf e}_{r}}{\rho_{*}} \frac{\partial p}{\partial r} 
\left( \rho_{1} + \frac{p_{1}}{c^{2}} +
\frac{2 \Gamma^{2} V \rho_{*}}{c^{2}} v_{1r} \right),
\label{eq:momentum2}
\end{eqnarray}
\begin{eqnarray}
\dot{p_{1}} + {\bf V} \nabla p_{1}
+ \frac{\gamma p}{\Gamma} \left[ \nabla \cdot ( \Gamma {\bf v}_{1}) + 
\frac{\Gamma^{3} V}{c^{2}} \left( \dot{v}_{1r} + 
V \frac{\partial v_{1r}}{\partial r} \right) \right] 
\nonumber \\
= - v_{1r} \left[ \frac{\partial p}{\partial r} + 
\frac{\gamma p V}{c^{2}} \frac{\partial}{\partial r} 
(V \Gamma^{2}) \right] - p_{1} \frac{\gamma}{r^{2} \Gamma} 
\frac{\partial}{\partial r} (r^{2} V \Gamma), 
\label{eq:energy2}
\end{eqnarray}
where $\rho_{*}= \rho + p/c^{2}$. In equations
(\ref{eq:continuity2})-(\ref{eq:energy2}) we took into account that
$\Gamma_{1}=\Gamma^{3} ({\bf V} \cdot {\bf v}_{1})/c^{2}$. Since a
quasi-steady state approximation is adopted, we have for the
unperturbed state $\Gamma^{2} \rho_{*} ({\bf V} \cdot \nabla ) {\bf V}
= - \nabla p$, and hence only the radial component of $\nabla p$ is
non-vanishing.

In a conical jet with a small opening angle, the characteristic
radial length scale of unperturbed quantities is of the order of the 
spherical radius and is much greater than the transverse length scale. 
Therefore, we can treat the radial and $\theta$-dependences of 
the perturbations separately. If we consider perturbations with 
the radial wave vector, $k$, satisfying the condition 
\begin{equation}
kr \gg 1,
\label{eq:kr}
\end{equation}
then the terms on the right hand sides of equations
(\ref{eq:continuity2})-(\ref{eq:energy2}), which contain only the
radial derivative of unperturbed quantities, are smaller by a factor
$kr$ than the corresponding terms on the left hand sides containing
the derivatives of perturbations.  Therefore, all terms on the
r.h.s. of equations (\ref{eq:continuity2})-(\ref{eq:energy2}) can be
neglected if inequality (\ref{eq:kr}) is fulfilled. The assumption
(\ref{eq:kr}) corresponds to the so--called local approximation in a
stability analysis. Under this assumption, we have from
(\ref{eq:continuity2})-(\ref{eq:energy2})
\begin{eqnarray}
\Gamma \left( \dot{\rho_{1}} + 
{\bf v}_{1} \cdot \nabla \rho + {\bf V} \cdot \nabla \rho_{1} \right) + 
\rho_{*}  \left[ \nabla ( \Gamma {\bf v}_{1}) + 
\right.
\nonumber \\
\left.
\frac{\Gamma^{3} V}{c^{2}} \left(\dot{v}_{1r}  
+ V \frac{\partial v_{1r}}{\partial r} \right) \right] = 0,
\label{eq:continuity3}
\end{eqnarray}
\begin{eqnarray}
\Gamma^{2} \rho_{*} \left[ \dot{\bf v}_{1} + ({\bf v}_{1} \cdot \nabla) 
{\bf V} + ({\bf V} \cdot \nabla) {\bf v}_{1} \right] + \nabla p_{1}
+ \frac{{\bf V}}{c^{2}} \dot{p}_{1} 
\nonumber \\
= 0,
\label{eq:momentum3}
\end{eqnarray}
\begin{eqnarray}
\dot{p_{1}} + {\bf V} \nabla p_{1}
+ \frac{\gamma p}{\Gamma} \left[ \nabla ( \Gamma {\bf v}_{1}) + 
\frac{\Gamma^{3} V}{c^{2}} \left( \dot{v}_{1r} + 
V \frac{\partial v_{1r}}{\partial r} \right) \right] 
\nonumber \\
= 0.
\label{eq:energy3}
\end{eqnarray}
Since the unperturbed quantities depend neither on $t$ nor on
$\varphi$ , the dependence of all perturbations on $t$, $r$ and
$\varphi$ can be taken in the form $\exp(i \omega t - i k r - i m
\varphi)$ if we adopt condition (\ref{eq:kr}). In the present
paper, we consider axisymmetric perturbations with $m=0$. Substituting
this dependence into equation (\ref{eq:energy3}), we can express the
perturbation of pressure in terms of ${\bf v}_{1}$,
\begin{equation}
p_{1} = \frac{i \gamma p}{\sigma \Gamma} \left[ \nabla \cdot 
(\Gamma {\bf v}_{1}) + i \sigma \frac{\Gamma^{3}}{c^{2}} V v_{1r} 
\right],
\label{eq:p1}
\end{equation}
where $\sigma = \omega - k V$. The quantity $kV$ is an advective
frequency and has a kinematic origin. The perturbations of the
velocity can be expressed in terms of $p_{1}$ from the momentum
equation (\ref{eq:momentum3}) which under our assumptions can be
rewritten as
\begin{equation}
\rho_{*} \Gamma^{2} \left( i \sigma 
{\bf v}_{1} + {\bf e}_{r} \frac{v_{1 \theta}}{r} 
\frac{\partial V}{\partial \theta} \right) = 
-i \left( \frac{\omega {\bf V}}{c^{2}} - i \nabla \right) p_{1}.
\label{eq:p1equ}
\end{equation}
Substituting expressions for $v_{1r}$ and $v_{1 \theta}$ into equation
(\ref{eq:p1}) and taking into account that for a small opening angle $\sin \theta 
\approx \theta$, we obtain the equation containing $p_{1}$ alone
\begin{eqnarray}
p_{1}'' +  \left[ \frac{1}{x} + \frac{2 \Gamma^{2} V'}{\sigma}
\left( k - \frac{\omega V}{c^{2}} \right) - \frac{\rho'}{\rho_{*}} 
\right] p_{1}' +
\nonumber \\
\Gamma^{2} \left[ \frac{\sigma^{2}}{c_{s}^{2}} - \left( k -
\frac{\omega V}{c^{2}} \right)^{2} \right] p_{1} = 0.
\label{eq:p1dif}
\end{eqnarray}
Here $c_{s} = \sqrt{\gamma p / \rho_{*}}$ is the sound speed, $x=r
\theta$, and the prime denotes a transverse derivative; for example,
$p_{1}' = d p_{1}/ d x$ where $dx=r d \theta$ . Equation
(\ref{eq:p1dif}) represents the behavior of small perturbations for
any velocity and density profiles within a conical jet with a small
opening angle. In our approach, the radial dependence of unperturbed
quantities enters in equation (\ref{eq:p1dif}) parametrically.

For our purposes, it will be convenient to use another form of the
equation (\ref{eq:p1dif}). Making the substitution
\begin{equation}
p_{1} = \sqrt{\frac{\rho_{*}}{x}} \exp \left[ - \int 
\frac{\Gamma^{2} V'}{\sigma}
\left( k - \frac{\omega V}{c^{2}} \right) dx \right] f,
\label{eq:p1_f}
\end{equation}
equation (\ref{eq:p1dif}) can be transformed into
\begin{equation}
f'' + q^{2}(r) f = 0,
\label{eq:f''}
\end{equation} 
where 
\begin{eqnarray}
q^{2} = \frac{\Gamma^{2} \sigma^{2}}{c_{s}^{2}}
- Q^{2} + \frac{2 \Gamma^{4} V'^{2}}{c^{2}} \left( \frac{1}{2}  
+ \frac{V}{c} \xi - \xi^{2} \right)
\nonumber \\
- \frac{\Gamma^{2}}{c} \left(\frac{V}{c} - \xi \right)
\left( \frac{\rho' V'}{\rho_{*}} - \frac{(xV')'}{x} \right), 
\label{eq:q2}
\end{eqnarray}
$\xi = kc /\sigma \Gamma^{2}$, and $Q^{2}$ is given by
\begin{equation}
Q^{2} = \frac{\sigma^{2} \Gamma^{2}}{c^{2}} \left( \frac{V}{c} - 
\xi \right)^{2} + 
\frac{3}{4} \left(\frac{\rho'}{\rho_{*}} \right)^{2} -
\frac{(x \rho')'}{2 x \rho_{*}} - \frac{1}{4 x^{2}}.
\end{equation}
Since in the present paper we consider a stratified sheared jet the
expression (\ref{eq:f''}) for $q^{2}$ differs from the corresponding
equation derived by Urpin (\cite{Urpin02}) only due to the presence of
terms containing derivatives of $\rho$.

In the core region where the temperature is high and the pressure is
determined by radiation, the sound speed, $c_{s}$, is of the order of
$c/\sqrt{3}$ and, hence, the acoustic frequency is comparable to that
of the light, $\sim kc$. Since the turnover time scale, $|V'|^{-1}$,
characterizing the rate of hydrodynamical processes associated with
shear is also very short, we can expect that typical frequencies in
our model are sufficiently large.  Therefore, we can try to obtain the
solution of equation (\ref{eq:f''}) in a high frequency limit when
$\xi \ll 1$, or
\begin{equation}
\mid \sigma \mid \gg k c / \Gamma^{2}.
\label{eq:sigmaapprox}
\end{equation}  
We will see that this inequality is well fulfilled in our jet model.
Restricting ourselves to the lowest terms in a small parameter $\xi$,
we obtain from the expression (\ref{eq:q2})
\begin{eqnarray}
q^{2} \approx \frac{\eta \sigma^{2} \Gamma^{2}}{c_{s}^{2}} 
+ \frac{\Gamma^{4} V'^{2}}{c^{2}} 
- \frac{V \Gamma^{2}}{c^{2}}
\left[ \frac{\rho' V'}{\rho_{*}} - \frac{(x V')'}{x} \right]
\nonumber \\
- \frac{3}{4} \left( \frac{\rho'}{\rho_{*}} \right)^{2}
+ \frac{(x \rho')'}{2 x \rho_{*}} + \frac{1}{4 x^{2}}, 
\label{eq:q2approx}
\end{eqnarray}
where $\eta = 1 - c_{s}^{2} V^{2}/c^{4}$. Since in the 
ultrarelativistic limit we have $c_{s}^{2} \approx c^{2}/3 $ and 
$V \rightarrow c$, the factor $\eta$ varies in a narrow range from 
$\sim 1$ to 2/3. 

The behavior of $q$ is determined by the $\theta$-dependence of
the jet velocity and density. We consider the case closely related 
to the calculations represented in Section 2. These simulations 
indicate that the density profile within the jet core can be fitted 
with a sufficient accuracy  by a simple exponential dependence,
\begin{equation}
\rho(r) \approx \rho_{0} e^{x/L_{\rho}},
\end{equation}
where $\rho_{0}$ is the density at the jet axis, and $L_{\rho}$ is the
characteristic length scale of stratification. Since the density
contrast between the surface and axis is of the order of 3-5 orders
of magnitude depending on the distance from the center of a collapsar, 
we can estimate $L_{\rho} \sim 0.1 x_{0}$. 

The velocity profile can be modeled by 
the parabolic dependence,
\begin{equation}
V(r) \approx V_{0} (1- x^{2}/L_{v}^{2}),
\end{equation}
where $V_{0}$ is the velocity at the jet axis, and $L_{v}$ is the
length scale of the velocity distribution, $L_{v} \sim x_{0}$.

Since $\Gamma^{2}$ is large in the jet core, the contribution of the
last four terms on the r.h.s. of equation (\ref{eq:q2approx}) which
are proportional to $\Gamma^{2}$ or $\Gamma^{0}$ is small compared to
the second term ($\propto \Gamma^{4}$) if we assume that $\Gamma^{2}
\gg \max (L_{v}/L_{\rho}, L_{v}/ \sqrt{x L_{\rho}})$ in the main
fraction of the core region where instability arises. After these
simplifications, we obtain the following expression for $q^{2}$ in the
jet core
\begin{equation}
q^{2} \approx \Gamma^{2} \left( \frac{\eta \sigma^{2}}{c_{s}^{2}} 
+ \frac{V'^{2} \Gamma^{2}}{c^{2}} \right). 
\label{eq:q2approx2}
\end{equation}

Consider the solution of equation (\ref{eq:f''}) by making use of a 
WKB-approximation which is well justified in those regions of the
jet where the condition  $|q x | > 1$ is fulfilled. With the accuracy 
in linear terms in a small parameter $|qx|^{-1}$, a WKB-solution of 
equation (\ref{eq:f''}) can be written as
\begin{equation}
f(x) = \frac{1}{\sqrt{q(x)}} \left[ C_{1} e^{i \int q d x'} 
+ C_{2} e^{-i \int q d x'}
\right],  
\label{eq:fWKB}
\end{equation} 
(Landau \& Lifshitz \cite{LL81}) where $C_{1}$ and $C_{2}$ are
constant that have to be chosen in such a way as to satisfy the
boundary conditions. In some cases, the condition $| q x | \gg 1$
holds everywhere within the jet, and the expression (\ref{eq:fWKB})
applies at $x_{0} \geq x \geq 0$. However, since the dependence of $q$
on $x$ is rather complex, it is generally also possible that the
condition of applicability can break at some point $x = x_{*} < x_{0}$
where $q \approx 0$. By analogy with quantum mechanics this point can
be called the turning point. We address this case as the most
interesting one from the point of view of applications to jets from
collapsars. Note that we have at $x=x_{*}$
\begin{equation}
\sigma^{2} \approx - \frac{1}{\eta} \left( \frac{c_{s}}{c} \right)^{2}
\Gamma^{2} V'^{2}   
\label{eq:sigma2}
\end{equation}
Since generally $\omega$ is complex, 
\begin{equation}
\omega \equiv \Omega - i / \tau,
\label{eq:omega}
\end{equation}
where $\Omega$ is the frequency of an unstable mode and $\tau$ is its
growth time, the condition (\ref{eq:sigma2}) can be satisfied only if
\begin{equation}
1/ \tau \gg | \Omega - k V|
\label{eq:1overtau}
\end{equation}
at the turning point. We are interested in the most powerful
instability (if it exists) with the growth time shorter than
$|kV|^{-1}$. For such instability the condition (\ref{eq:1overtau})
should be fulfilled everywhere within the jet. If equation
(\ref{eq:1overtau}) holds then $q^{2}$ is approximately real and
$q^{2} < 0$ at $x_{*} > x \geq 0$ and $q^{2} > 0$ at $x_{0} > x >
x_{*}$.

It is convenient to represent a WKB-solution from both sides of the
turning point in the form
\begin{equation}
f(x) = \frac{1}{\sqrt{q(x)}} \left[ C_{1}^{(r,l)} e^{i \int_{x_{*}}^{x} 
q d x'} + C_{2}^{(r,l)} e^{-i \int_{x_{*}}^{x} q d x'}
\right],  
\label{eq:fWKB2}
\end{equation} 
(Landau \& Lifshitz \cite{LL81}) where $C_{1}^{(r,l)}$ and
$C_{2}^{(r,l)}$ are constant; indexes $r$ and $l$ refer to the right
($x > x_{*}$) and left ($x < x_{*}$) from $x= x_{*}$,
respectively. Note that generally $C_{1}^{(r)} \neq C_{1}^{(l)}$ and
$C_{2}^{(r)} \neq C_{2}^{(l)}$.

To obtain the eigenvalues and eigenfunctions of equation
(\ref{eq:f''}) one needs the boundary conditions for
perturbations. One of the boundary conditions is obvious: since
$p_{1}$ is a physical quantity it should be finite everywhere including
the jet axis and, hence, $f$ should vanish at $x=0$. If the turning
point is not very close to 0, then usually it is sufficient to choose
an exponentially decreasing solution at $x < x_{*}$. It is known (see,
e.g., Landau \& Lifshitz \cite{LL81}) that the solution
(\ref{eq:fWKB2}) at $x > x_{*}$ matches an exponentially decreasing
solution beyond the turning point (at $x < x_{*}$) if $C_{1}^{(l)}=
(C/2) \exp (i \pi / 4)$ and $C_{2}^{(l)}= (C/2) \exp (- i \pi / 4)$
where $C$ is constant. Hence, the solution at $x>x_{*}$ satisfying the
true boundary condition at $x=0$ has the form
\begin{equation}
f(x) = \frac{C}{\sqrt{q(x)}} \cos \left( \int_{x_{*}}^{x} q(x') d x'
+ \frac{\pi}{4} \right).
\end{equation}
 
The formulation of a true boundary condition at the outer boundary can
face some problems but, fortunately, the properties of modes with a
relatively small length scale in the $\theta$-direction are not
sensitive to the particular choice of boundary conditions. We consider
the case when the density in a surrounding medium is much larger than
that in the jet. In this case (see, e.g., Glatzel \cite{Glatzel88}, Wu
\& Wang \cite{WW91}), a much denser surrounding medium plays the role
of a wall and the perturbations of the displacement in the
$\theta$-direction vanish at the outer boundary, $x= x_{0}$. As
it follows from equation (\ref{eq:p1equ}), $v_{1 \theta}$ vanishes if
$d p_{1}/dx =0$ at $x=x_{0}$.  Using the expression (\ref{eq:p1_f}),
we can rewrite this condition as
\begin{equation}
f' + \left[ \frac{\Gamma'}{\Gamma} \left( 1 - \frac{c}{V} \xi \right) 
+ \frac{\rho'}{2 \rho_{*}} - 
\frac{1}{2x_{0}} \right] f = 0.
\end{equation}
Since we consider the modes satisfying condition
(\ref{eq:sigmaapprox}) and, in a WKB-approximation, the wavelength of
perturbations is assumed to be much shorter than the characteristic
length scales of any unperturbed quantity, the boundary condition at
$x=x_{0}$ can approximately be represented as
\begin{equation}
f'(x_{0}) \approx 0.
\end{equation}
With this boundary conditions, the dispersion equation reads
\begin{equation}
\int_{x_{*}}^{x_{0}} q(x) dx = \pi \alpha, 
\label{eq:dispersion}
\end{equation}
where $\alpha = n + 3/4$, $n$ is integer. Substituting the expression
(\ref{eq:q2approx2}) for $q$ in (\ref{eq:dispersion}), we have
\begin{equation}
\int_{x_{*}}^{x_{0}} \Gamma \left( \frac{\eta \sigma^{2}}{c_{s}^{2}} 
+ \frac{V'^{2} \Gamma^{2}}{c^{2}} \right)^{1/2}
dx = \pi \alpha. 
\label{eq:dispersion2}
\end{equation}

We can estimate the characteristic roots of equation
(\ref{eq:dispersion2}) using the mean value theorem. We have
\begin{equation}
\Gamma_{a} \left( \frac{\eta_{a} \sigma^{2}_{a}}{c_{sa}^{2}} 
+ \frac{V_{a}'^{2} \Gamma_{a}^{2}}{c^{2}} \right)^{1/2}
= \frac{\alpha \pi}{\Delta x},
\end{equation}
where $\Delta x = x_{0}-x_{*}$ and the subscript $a$ means the value
of a quantity at $x=x_{a}$, $x_{a}$ is a mean point within the range
from $x_{*}$ to $x_{0}$. Since under the condition (\ref{eq:1overtau})
we have $\sigma_{a}^{2} \approx - 1/ \tau^{2}$, then the growth rate
is given by
\begin{equation}
\frac{1}{\tau^{2}} \approx \frac{c_{sa}^{2} V_{a}'^{2} 
\Gamma_{a}^{2}}{\eta_{a} c^{2}} \left( 1 -
\frac{\alpha^{2}}{N^{2}} \right),
\end{equation} 
where 
\begin{equation}
N^{2} = \frac{(\Delta x)^{2} V_{a}'^{2}}{\pi^{2} c^{2}} 
\; \Gamma_{a}^{4}. 
\end{equation}
If $n < N -3/4$, then $\tau^{2}$ is positive and perturbations 
with such n are unstable. On the contrary, perturbations with very
large $n$ satisfying the condition $n > N - 3/4$ are stable. Note 
that $N \sim \Gamma_{a}^{2}$. In the case $n \ll N$, we have
\begin{equation}
\frac{1}{\tau} \approx \frac{1}{\sqrt{\eta_{a}}} \; \frac{c_{sa}}{c}
\; \Gamma_{a} |V_{a}'|.
\end{equation}

At $c_{s} \sim c/\sqrt{3}$, we obtain a very simple estimate of the 
growth time
\begin{equation}
\frac{1}{\tau} \sim \Gamma_{a} |V_{a}'|.
\label{eq:1overtauestimate}
\end{equation}
Jets with a relativistic sound speed turn out to be unstable for any
velocity profile. The growth time is typically very short and depends
on the shear being shorter for larger $|V'|$. Also, the instability
arises faster for jets with a very high Lorentz factor.  Note,
however, that the origin of the factor $\Gamma_{a}$ in equation
(\ref{eq:1overtauestimate}) is purely kinematic. The growth time of
the instability in the comoving frame is of the order of $|V'|$, but
since we consider the instability in the rest frame, the Lorentz
transformation of the ``comoving'' growth time yields the result
(\ref{eq:1overtauestimate}).

The characteristic growth length, $L=V \tau$, can be estimated, for
$c_{s} \sim c/ \sqrt{3}$, as
\begin{equation}
L \sim \frac{1}{\Gamma_{a}} \frac{V}{|V'_{a}|}
\sim r \frac{\theta_{0}}{\Gamma_{a}}, 
\end{equation}
and is much shorter than the radius of a collapsar. Therefore, the
instability can manifest itself even at the very early stage of the
evolution of the jet.

\section{Discussion}

We have treated numerically and analytically the instability that can
arise in jets from collapsars. The instability is caused by a combined
action of shear, which is unavoidable in such jets, and an extremely
high compressibility associated with a relativistic sound speed. It
turns out that jets from collapsars are more unstable than, for
example, standard extragalactic jets because of the relativistic
compressibility. The fact of instability itself does not depend on a
particular shape of the velocity profile: the instability can arise
for any dependence $V(\theta)$. However, the growth time is shorter
for flows with a stronger shear. Note that only non-homogeneous
perturbations in the radial direction ($k \neq 0$) can be unstable.

{\bf 
Although it is commonly believed that three dimensional studies of the
stability of relativistic jets will include additional instable modes
(in many cases even more unstable that the axisymmetric ones), the
recent work of Hardee \& Rosen \cite{HR02}, has pointed the fact that shear
leads both to an enhancement of the axisymmetric modes and a
suppression of the asymmetric modes. Hence, a fully three dimensional
study would not yield to radically different results from the ones
that we obtain here.}

\begin{figure*}
\epsfxsize=15cm
\epsffile{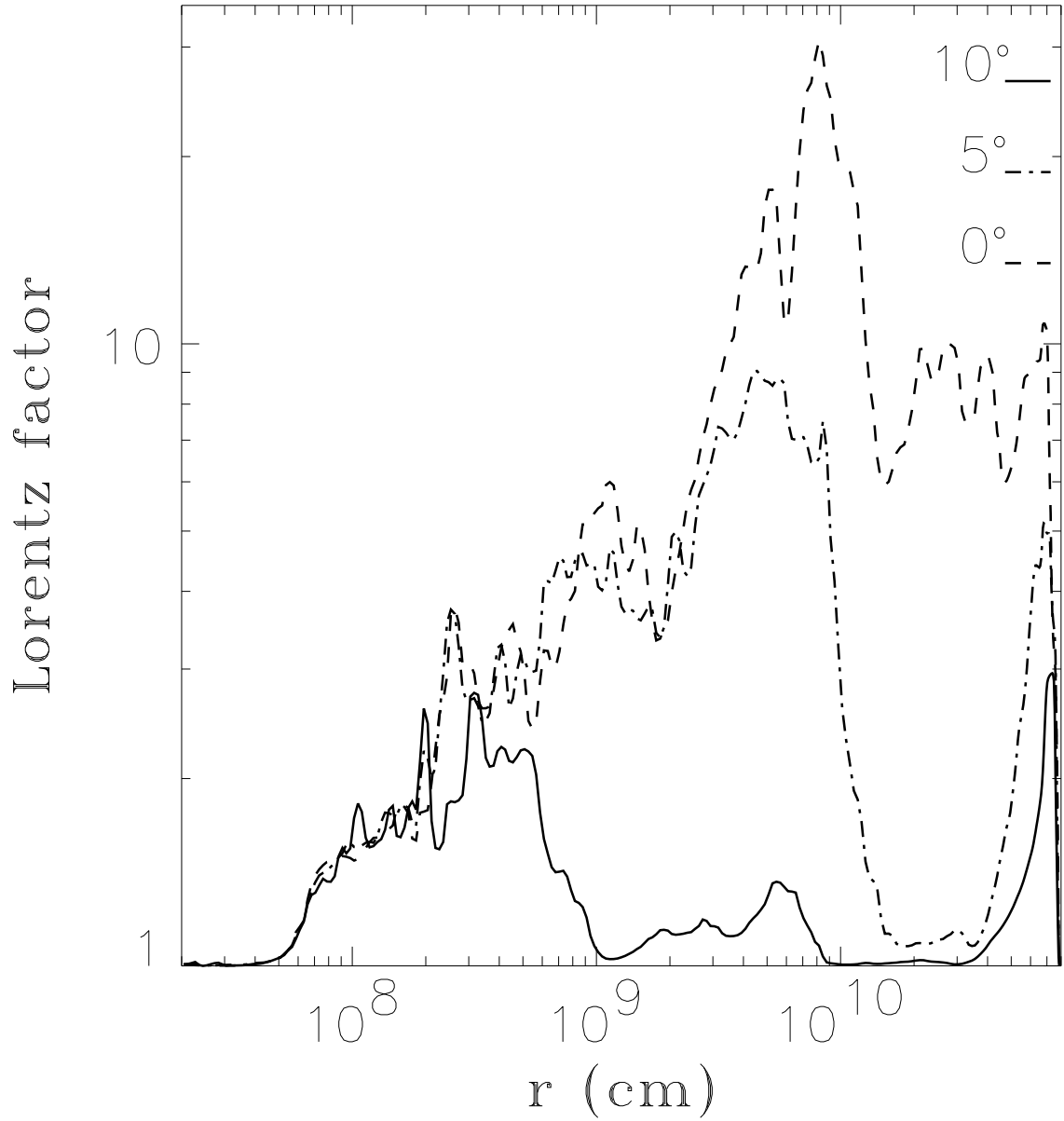}
%
\caption{Lorentz factor as a function of the radius for three
different angles ($\theta\approx 0^{\circ}, 5^{\circ}, 10^{\circ}$;
see legends in the figure) after an evolution time of $5.24\,$s.}
\label{fig:lorentz}
\end{figure*}

  In the main fraction of the jet volume, the instability grows very
rapidly. If we estimate the average Lorentz factor as $\sim 10$ and
the half-opening angle as $\sim 5^{\circ}$ then the growth time
(\ref{eq:1overtauestimate}) is of the order $0.01 r/c$. This is much
shorter than the life time of the jet even at $r \sim R$. Therefore,
we can expect that shortly after jet formation the instability will
generate well developed turbulent motions with substantial
fluctuations of the Lorentz factor, density, pressure, etc. This
conclusion is in good agreement with the results of numerical
simulations (see Figs.~\ref{fig:lorentz} and
\ref{fig:density}). During the jet's propagation, fluctuations can
become noticeable at a relatively early evolutionary stage and, hence,
at a small distance from the formation region because the growth time
is sufficiently short even in the inner region of the collapsar. Of
course, the initial amplitude of fluctuations in the jet is quite
uncertain.  In our case, the initial fluctuations are triggered by
numerical reasons, but it is quite likely that they mimic the
irregularities of the process of accretion onto the central black
hole.  Usually the time $\tau_{0}$ required for an instability to
amplify the amplitude of perturbations to a noticeable value is longer
than the growth time, $\tau$, by some factor which is typically $\sim
(5-10)$, so $\tau_{0} \sim (5-10) \tau$. After this time, the
fluctuations have grown by a factor $\sim 10^{2}-10^{4}$ compared to
their initial value which probably is sufficient to become
significant. In the region where $\tau_{0}$ is shorter than the
propagation time scale, which can be estimated as $\sim r/c$,
fluctuations reach noticeable values. On the contrary, fluctuations
seem to be insignificant in regions where $\tau_{0} > r/c$. We can
define the radius where the instability starts to manifest itself as
that where the condition $\tau_{0} \sim r/c$ is
fulfilled. Substituting $\tau_{0}$, we obtain that
\begin{equation}
(5-10) \frac{\theta_{0}}{\Gamma_{a}} \sim 1 \xi
\end{equation}
at this radius. At small $r$ ($\sim 8 \times 10^{8}\,$cm well below
jet break-out), the jet is less collimated ($\theta_{0} \approx
10-12^{\circ}$) than on average (see Aloy et
al. \cite{Aetal00}). Therefore, the instability should manifest 
already in regions where $\Gamma \sim 1-2$.  This conclusion is in
qualitative agreement with what is shown in Fig.~\ref{fig:lorentz}
where appreciable fluctuations appear in regions with a Lorentz factor
between 1 and 2.
 
Note that the radial wave vector of unstable perturbations, $k$,
should satisfy rather restrictive conditions. First, $k$ has to be
sufficiently large for the applicability of the local approximation
(see equation (\ref{eq:kr})).  On the other hand, condition
(\ref{eq:1overtau}) which is necessary for the existence of the
turning point, implies that
\begin{equation}
\Gamma_{a} \frac{|V'_{a}|}{V} \sim \frac{\Gamma_{a}}{x_{0}} \gg k.
\label{eq:GV'ggk} 
\end{equation}
If the inequality (\ref{eq:GV'ggk}) holds then the condition
(\ref{eq:sigmaapprox}) used in our calculations is certainly fulfilled
and the parameter $\xi$ is small.  Combining equations (\ref{eq:kr})
and (\ref{eq:GV'ggk}) and taking into account that $x_{0}= r
\theta_{0}$, we obtain that
\begin{equation}
\frac{\Gamma_{a}}{\theta_{0}} \gg kr \gg 1 
\end{equation}
for unstable perturbations. Estimating $\Gamma \sim 10$ and
$\theta_{0} \sim 5-6^{\circ}$ in the main fraction of the jet volume,
we have $100 \gg kr \gg 1$. Numerical simulations (see
Fig.~\ref{fig:lorentz}) indicate that the characteristic length scale
of fluctuations depends on $r$ and varies within the range from $0.1r$
to $0.5r$ in good agreement with the prediction of the theory. Note
that fluctuations with a shorter length scale cannot be resolved with
the computational grid used in the simulations.

All this allows one to speculate that the calculated fluctuations of
parameters within the jet are physical (i.e., not simply numerical
artifacts) and reflect the presence of a very strong shear-driven
instability. We can expect that inhomogeneities caused by this
instability will produce shocks in the course of their non-linear
evolution when faster fluctuations try to overtake slower ones or when
fluctuations moving in the positive and negative radial directions
collide.  If this is the case then internal shocks might be more or
less randomly distributed and oriented within the jet forming
filamentary structures. It is often supposed that shocks in a
ultrarelativistic wind or jet are responsible for GRBs themselves
whereas the impact against the ambient matter of this wind produces an
external shock which likely produces the observed afterglows (Rees \&
M\'esz\'aros \cite{RM92}, \cite{RM94}).  Shocks can convert a portion
of kinetic energy into a non-thermal gamma/X-ray transient emission
which is usually ascribed to particle acceleration by
shocks. Typically, the efficiency of this conversion is not high,
$\sim 1-2$ \%, but it can be much greater ($\sim 20-40$ \%) for the
interaction of fluctuations with very different Lorentz factors
(Kobayashi, Piran \& Sari \cite{KPS97}, Kobayashi \& Sari
\cite{KS01}). Since in our model the calculated fluctuations move with
substantially different Lorentz factors we can expect a highly
efficient transformation of their kinetic energy into radiation.  The
proposed instability can also accounts for the rapid variability of
the gamma-ray light curves, which lasting from tens to hundreds of
seconds, exhibit variability sometimes down to milliseconds (Fishman
\& Meegan \cite{FM95}). Likely, the most rapid temporal variability
associated with the shear-driven instability has a time scale of the
order of the growth time (\ref{eq:1overtauestimate}). At the surface
of a collapsar, for example, this time scale is as short as $\sim
10^{-3}$s. Since the jet is highly inhomogeneous and the Lorentz
factor varies strongly during the jet's propagation, a slower
variability could also be represented in the gamma-ray light curves.

Another remarkable inference from the considered model is that the
turbulent motions caused by the instability may also be important for
the electron-proton energy exchange and, particularly, for the
generation of the magnetic field in jets from collapsars.

\section*{Acknowledgement}

This research was supported in part by the Russian Foundation of Basic
Research and Deutsche Forschungsgemeinschaft (grant
00-02-04011). V.U. thanks Ministerio de Educaci\'on, Cultura y Deporte
of Spain for the financial support under the grant
SAB1999-0222. M.A.A. acknowledges the EU-Commission for a fellowship
(MCFI-2000-00504). M.A.A. thanks E. M\"uller and P.E. Hardee for their
enlightening discussions.


\end{document}